\documentstyle[12pt,a4]{article} 

\begin{document} 

\reversemarginpar

\title{On the response of detectors in classical\\ 
electromagnetic backgrounds}
\author{L.~Sriramkumar\thanks{E-mail:~sriram@racah.phys.huji.ac.il}\\ 
Racah Institute of Physics, Hebrew University\\ 
Givat Ram, Jerusalem~91904, Israel}

\date{}
\maketitle
\begin{abstract}
I study the response of a detector that is coupled 
non-linearly to a quantized complex scalar field in 
different types of classical electromagnetic backgrounds.
Assuming that the quantum field is in the vacuum state, 
I show that, when in {\it inertial}\/ motion, the 
detector responds {\it only}\/ when the electromagnetic 
background produces particles.  
However, I find that the response of the detector is
{\it not}\/ proportional to the number of particles 
produced by the background.
\end{abstract}
\newpage

\section{The concept of a detector}

The idea of the detector originated in literature with 
the aim of providing an operational definition for the 
concept of a particle and also for being utilized as a 
probe to study the phenomenon of particle production in 
classical gravitational backgrounds.
A detector is an idealized point like object whose motion is 
described by a classical worldline, but which nevertheless 
possesses internal energy levels.
Such detectors are essentially described by the interaction 
Lagrangian for the coupling between the degrees of freedom 
of the detector and the quantum field.
The response of detectors that are coupled to the quantum 
field through a linear~\cite{unruh76,dewitt79} or a derivative 
coupling~\cite{hinton83,hinton84} and also detectors that 
are coupled to the energy-momentum tensor of the quantum 
field~\cite{paddytp87} have been studied in a variety of 
situations in different classical gravitational backgrounds 
(also see Ref.~\cite{bd82}, Secs.~3.3, 5.4 and~8.3 in this 
context).

Phenomena such as vacuum polarization and particle production
that occur in gravitational backgrounds take place in 
electromagnetic backgrounds too.
It will be interesting to examine as to how detectors respond 
to such phenomena in classical electromagnetic backgrounds.
But, the response of detectors on non-inertial trajectories 
turns out to be non-zero even in the Minkowski vacuum in flat 
spacetime~\cite{letaw81,paddy82}.
Therefore, in order to avoid effects due to non-inertial 
motion and also to isolate the effects that arise due to
the electromagnetic background, it is essential that we
restrict our attention to inertial trajectories.  
With this motivation, in this Letter, I shall study the 
response of an {\it inertial}\/ detector in different 
types of classical electromagnetic backgrounds.
(I shall set $\hbar=c=1$ and I shall denote complex and 
Hermitian conjugation by an asterisk and a dagger, respectively.)

\section{The non-linearly coupled detector}

The quantum field I shall consider is a {\it complex}\/ 
scalar field~$\Phi$ described by the action
\begin{equation}
S[\Phi]=\int d^4x\, \left[(D_{\mu}\Phi)(D^{\mu}\Phi)^*
- m^2\Phi\Phi^*\right],\label{eqn:emact}
\end{equation}
where $D_{\mu}\equiv\left(\partial_{\mu}+iq A_{\mu}\right)$, 
$A^{\mu}$ is the vector potential describing the classical 
electromagnetic background and $q$ and $m$ are the charge 
and the mass of a single quanta of the scalar field.
Varying this action leads to the following equation of motion
for the complex scalar field~$\Phi$:
\begin{equation}
\left(D_{\mu}D^{\mu} +m^2\right)\Phi=0.\label{eqn:kg}
\end{equation}

The simplest of the different possible detectors is the 
detector due to Unruh and DeWitt~\cite{unruh76,dewitt79}. 
Consider a Unruh-DeWitt detector that is moving along a 
trajectory ${\tilde x}(\tau)$, where ${\tilde x}$ denotes 
the set of four coordinates $x^{\mu}$ and $\tau$ is the 
proper time in the frame of the detector.
The interaction of the Unruh-DeWitt detector with a 
{\it real}\/ scalar field~$\Phi$ is described by the 
Lagrangian
\begin{equation}
{\cal L}_{\rm int}
=c\, \mu(\tau)\, \Phi\left[{\tilde x}(\tau)\right],
\label{eqn:lint}
\end{equation}
where $c$ is a small coupling constant and $\mu(\tau)$ 
is the detector's monopole moment.
For the case of the complex scalar field I shall be 
considering here, this interaction Lagrangian can be 
generalized to
\begin{equation}
{\cal L}_{\rm int}
=c\,\biggl(\mu(\tau)\, \Phi[{\tilde x}(\tau)]
+ \mu^*(\tau)\, \Phi^*[{\tilde x}(\tau)]\biggl).
\label{eqn:lintc}
\end{equation}
But, under a gauge transformation of the form: $A^{\mu} \to 
\left(A^{\mu}+\partial^{\mu}\chi\right)$, the complex scalar 
field transforms as: $\Phi\to \left(\Phi\, e^{-iq\chi}\right)$. 
Clearly, the interaction Lagrangian~(\ref{eqn:lintc}) will 
{\it not}\/ be invariant under such a gauge transformation, 
unless I assume that the monopole moment transforms as 
follows: $\mu\to \left(\mu\, e^{iq\chi}\right)$. 
However, I would like to treat the detector part of 
the coupling, viz. the monopole moment $\mu(\tau)$, 
as a quantity that transforms as a scalar under gauge 
transformations. 
In such a case, the simplest of the Lagrangians that is 
{\it explicitly}\/ gauge invariant is the {\it non-linear}\/ 
interaction
\begin{equation}
{\cal L}_{\rm int} 
= c\, \mu(\tau)\, \biggl(\Phi[{\tilde x}(\tau)]\,
\Phi^*[{\tilde x}(\tau)]\biggl).\label{eqn:giint}
\end{equation}
In what follows, I shall study the response of a detector 
that is coupled to the field through such an interaction 
Lagrangian in different types of classical electromagnetic 
backgrounds.
It is important to note here that demanding gauge invariance 
naturally leads to non-linear interactions.
A physical manifestation of gauge invariance is charge 
conservation.  
As we shall see later, the non-linear and gauge invariant 
interaction Lagrangian~(\ref{eqn:giint}) leads to the 
excitation of a particle-anti-particle pair thereby conserving 
charge.

In an electromagnetic background, the quantized complex 
scalar field~${\hat \Phi}$ satisfying the Klein-Gordon 
equation~(\ref{eqn:kg}) can, in general, be decomposed 
as follows (see Ref.~\cite{fulling89} and references
therein):
\begin{equation}
{\hat \Phi}({\tilde x})
=\sum_i \left[{\hat a}_i\, u_i({\tilde x}) 
+{\hat b}_{i}^{\dag}\, v_{i}({\tilde x})\right],
\label{eqn:decompc}
\end{equation}
where $u_{i}({\tilde x})$ and $v_{i}({\tilde x})$ are positive 
and negative {\it norm}\/ modes, respectively\footnote{The only 
non-trivial commutation relations satisfied by the two sets of 
operators $\left\{{\hat a}_{i},{\hat a}_{i}^{\dag}\right\}$ and 
$\left\{{\hat b}_{i},{\hat b}_{i}^{\dag}\right\}$ are: 
$\left[{\hat a}_{i},{\hat a}_{j}^{\dag}\right]
=\left[{\hat b}_{i}, {\hat b}_{j}^{\dag}\right]
=\delta_{ij}$. 
All other commutators vanish.}.
These modes are normalized with respect to the following 
gauge invariant scalar product (see, for e.g., 
Ref.~\cite{fulling89})
\begin{equation}
(u_{i},u_{j})
=-i\int\limits_{t=0}^{} d^3x\, \left(u_{i}
\left[\partial_t-iqA_t\right]u_{j}^*
-u_{j}^* \left[\partial_t+iqA_t\right]u_{i}\right),
\label{eqn:sclrprdct}
\end{equation}
where $\partial_t\equiv(\partial/\partial t)$ and~$A_t$ 
is the zeroth component of the vector potential $A^{\mu}$.
The vacuum state~$\vert 0\rangle$ of the quantum 
field~${\hat \Phi}$ is defined as the state that 
is annihilated by {\it both}\/ the 
operators~${\hat a}_{i}$ and ${\hat b}_{i}$ for 
{\it all}\/ $i$.

Now, assume that the quantized complex scalar field~${\hat \Phi}$ 
is initially in the vacuum state~$\vert 0\rangle$. 
Then, up to the first order in perturbation theory, 
the amplitude of transition of the detector that is 
coupled to the field through the interaction 
Lagrangian~(\ref{eqn:giint}) is given by
\begin{equation}
{\cal A}({\cal E}) 
=\left(\frac{{\cal M}}{2}\right)\,
\int\limits_{-\infty}^{\infty} d\tau\, e^{i {\cal E}\tau}\, 
\langle\Psi\vert\biggl({\hat \Phi}[{\tilde x}(\tau)]\,
{\hat \Phi}^{\dag}[{\tilde x}(\tau)]
+ \,{\hat \Phi}^{\dag}[{\tilde x}(\tau)]\,
{\hat \Phi}[{\tilde x}(\tau)]\biggl)\vert 0\rangle,
\label{eqn:detampnld}
\end{equation}
where ${\cal M}\equiv i c\,\langle E\vert {\hat \mu}(0)
\vert E_{0}\rangle$, ${\cal E}=(E-E_0)$, $E_0$ and $E$ 
are the energy eigen values corresponding to the ground 
state $\vert E_0\rangle$ and the excited state $\vert 
E\rangle$ of the detector and $\vert \Psi\rangle$ is the 
state of the quantum field after its interaction with the 
detector.
(Since the term~${\cal M}$ depends only on the internal 
structure of the detector and not on its motion, I shall 
drop this term hereafter.)
The transition amplitude~${\cal A}({\cal E})$ above involves 
products of the field~${\hat \Phi}$ at the {\it same}\/ 
spacetime point and hence we will encounter divergences when 
evaluating this transition amplitude.
In order to avoid the divergences, I shall normal order 
the creation and the annihilation operators in the matrix 
element in the transition amplitude~(\ref{eqn:detampnld}).
On substituting the decomposition~(\ref{eqn:decompc}) 
for the field~${\hat \Phi}$ in the transition 
amplitude~(\ref{eqn:detampnld}) and normal ordering the
creation and the annihilation operators, I obtain that
\begin{equation}
{\cal A}^*({\cal E})
=\sum_{i}\sum_{j}
\int\limits_{-\infty}^{\infty} d\tau\, e^{-i {\cal E}\tau}\, 
u_{i}[{\tilde x}(\tau)]\,  v_{j}^*[{\tilde x}(\tau)]\; 
\langle 0\vert {\hat a}_{i}{\hat b}_{j}
\vert\Psi\rangle.\label{eqn:transamp1}
\end{equation}
This transition amplitude will be non-zero {\it only}\/ when 
$\vert \Psi\rangle= {\hat a}_{i}^{\dag}{\hat b}_{j}^{\dag}\vert 
0\rangle =\vert 1_{i}, 1_{j}\rangle$.
This implies that the interaction of the field with the 
detector leads to the excitation of a particle-anti-particle 
pair. 
Since the quantum field I am considering here is a 
{\it charged}\/ scalar field, the excitation of a 
particle-anti-particle pair is essential for charge 
conservation.
As I had pointed out before, it is the non-linear 
and the gauge invariant nature of the interaction 
Lagrangian~(\ref{eqn:giint}) that ensures that such 
a pair is indeed excited.

Before I go on to study the response of inertial detectors 
in electromagnetic backgrounds, let me briefly discuss the 
response of an inertial detector in the Minkowski vacuum.
(The arguments I shall present here will prove to be useful 
for our discussion later on.)
Consider an inertial detector stationed at a point, say,~${\bf a}$.
In the absence of an electromagnetic background, the
positive and negative norm modes are related as follows: 
$v_{i}({\tilde x})=u_{i}^*({\tilde x})$.
Moreover, in the Minkowski coordinates, the definition of 
positive norm modes {\it match}\/ the definition of positive 
frequency modes.
Then, it is clear from Eq.~(\ref{eqn:transamp1}) that it is 
{\it only}\/ the positive frequency modes $u_i({\tilde x})$ 
that contribute to the transition 
amplitude~${\cal A}^*({\cal E})$ in such a situation.
Therefore, the transition amplitude of the detector
corresponding to a pair of modes, say, ${\bf k}$ 
and ${\bf l}$, of the quantum field is given by
\begin{equation}
{\cal A}^*({\cal E})
=\left(\frac{e^{i({\bf k}+{\bf l}).{\bf a}}}{\sqrt{(2\pi)^4\, 
4\omega_{k}\omega_{l}}}\right)\; \delta^{(1)}({\cal E}+
\omega_{k}+\omega_{l}),\label{eqn:minkampnld}
\end{equation}
where, for a given mode ${\bf k}$,
$\omega_{k}=\left(\vert {\bf k}\vert^2+m^2\right)^{1/2}$.
The quantities $\omega_{k}$ and $\omega_{l}$ are always 
$\ge m$ and, since ${\cal E}>0$ as well, the argument of 
the delta function above is a positive definite quantity 
and, hence, the transition amplitude ${\cal A}^*({\cal E})$ 
reduces to zero for {\it all}\/ ${\bf k}$ and ${\bf l}$. 
In other words, the non-linearly coupled detector will 
{\it not}\/ respond in the Minkowski vacuum state when 
in inertial motion.

The transition probability of the non-linearly coupled
detector to all possible final states $\vert \Psi\rangle$ 
of the field can now be evaluated from the transition 
amplitude~(\ref{eqn:transamp1}).
I find that
\begin{equation}
{\cal P}({\cal E}) 
=\sum_{\vert \Psi\rangle} 
\vert{\cal A}({\cal E})\vert^2
= \int\limits_{-\infty}^\infty d\tau\, 
\int\limits_{-\infty}^\infty d\tau'\, 
e^{-i{\cal E}(\tau-\tau')}\, 
{\tilde G}\left[{\tilde x}(\tau), {\tilde x}(\tau')\right],
\label{eqn:detprobnld}
\end{equation}
where ${\tilde G}\left[{\tilde x}(\tau), 
{\tilde x}(\tau')\right]$ 
is a four point function given by
\begin{equation}
{\tilde G}[{\tilde x},{\tilde x'}]
=\sum_{i} \left[u_i({\tilde x})u_i^*({\tilde x'})\right]\;
\sum_j \left[v_j^*({\tilde x})\, v_j({\tilde x'})\right].
\label{eqn:fptfn}
\end{equation}
In cases wherein the four point function 
${\tilde G}[{\tilde x},{\tilde x'}]$ is invariant under 
translations in the proper time in the frame of the detector, 
a transition probability rate for the detector can be defined 
as follows:
\begin{equation}
{\cal R}({\cal E})
=\int\limits_{-\infty}^{\infty}d(\tau-\tau')\,
e^{-i{\cal E}(\tau-\tau')}\,
{\tilde G}(\tau-\tau').\label{eqn:detrate}
\end{equation} 

I had pointed out above that, in the absence of an 
electromagnetic background, the positive and negative
norm modes are related by the following expression:
$v_{i}({\tilde x})=u_{i}^*({\tilde x})$.
It is then useful to note that, in such a case, the four 
point function~${\tilde G}[{\tilde x},{\tilde x'}]$ is given 
by square of the Wightman function in the Minkowski vacuum. 
Therefore, when in inertial motion, the transition probability 
rate~${\cal R}({\cal E})$ of the non-linearly coupled detector 
in the Minkowski vacuum is identically zero (for exactly the 
same reasons) as it is in the case of the Unruh-DeWitt detector 
(see Ref.~\cite{bd82}, pp.~50--53 in this context).

In the following three sections, I shall study the response 
of the non-linearly coupled detector (when it is in 
{\it inertial}\/ motion) in:~(i)~a time-dependent electric 
field, (ii)~a time-independent electric field and 
(iii)~a time-independent magnetic field, backgrounds. 
I shall then conclude this Letter with a few summarizing 
remarks.

\section{In time-dependent electric field backgrounds}

Consider a time-dependent electric field background described 
by vector potential
\begin{equation}
A^{\mu}=(0, A(t), 0,0),\label{eqn:tde}
\end{equation}
where $A(t)$ is an arbitrary function of $t$.
This vector potential gives rise to the electric field 
${\bf E}=-(dA/dt)\, {\hat {\bf x}}$, where ${\hat {\bf x}}$ 
is the unit vector along the positive $x$-direction.
The modes of a quantum field evolving in such a time-dependent 
electric field background are of the form
\begin{equation}
u_{\bf k}(t,{\bf x})
=g_{\bf k}(t)\; e^{i{\bf k}\cdot{\bf x}}.
\label{eqn:modetde}
\end{equation}
In general, the modes at early and late times will be 
related by a non-zero Bogolubov coefficient $\beta$  
(see Ref.~\cite{mp77} and references therein).
In fact, the expectation value of the number operator 
(corresponding to a given mode of the quantum field) 
at late times in the in-vacuum will be proportional 
to $\vert \beta \vert^2$.

Now, consider a detector that is stationed at a particular
point.
Along the world line of such a detector, the four point
function~(\ref{eqn:fptfn}) corresponding to the 
modes~(\ref{eqn:modetde}) is given by
\begin{equation}
{\tilde G}(t,t')
=\sum_{\bf k} \sum_{\bf l} \left[g_{\bf k}(t)\, 
g_{\bf l}(t)\, g_{\bf k}^*(t')\, g_{\bf l}^*(t')\right]
\end{equation}
and the transition probability of the detector reduces to
\begin{equation}
{\cal P}({\cal E})
=\sum_{\bf k}\sum_{\bf l} \vert g_{\bf kl}({\cal E})\vert^2,
\quad {\rm where}\quad
g_{\bf kl}({\cal E})
=\int\limits_{-\infty}^{\infty}dt\, e^{-i{\cal E}t}\, 
\left[g_{\bf k}(t)\, g_{\bf l}(t)\right].
\end{equation}
Clearly, the response of the inertial detector will, in
general, be non-zero.

Let me now assume that the function $A(t)$ behaves such 
that the electric field vanishes in the past and future 
infinity.
Also, let the detector be switched on for a finite time 
interval in the future asymptotic domain.
Let me further assume that the effects that arise due 
to switching~\cite{ftdet1,ftdet2,ftdet3} can be neglected.
Then, by relating the modes at future and past infinity, 
I can express the transition probability rate of the detector 
(in the in-vacuum) in terms of the Bogolubov coefficients 
$\alpha$ and $\beta$ as follows:
\begin{eqnarray}
{\cal R}({\cal E})
&=&(2\pi)\,\sum_{\bf k}\sum_{\bf l}
\biggl(2\, \vert \alpha_{\bf k}\vert^2\;
\vert \beta_{\bf l}\vert^2\;
\delta^{(1)}({\cal E}+\omega_{k}-\omega_{l})\nonumber\\
& &\qquad\qquad\qquad\qquad\quad
+\, \vert \beta_{\bf k}\vert^2\; 
\vert \beta_{\bf l}\vert^2\;
\delta^{(1)}({\cal E}-\omega_{k}-\omega_{l})\biggl),
\end{eqnarray}
where $\omega_{k}$ and $\omega_{l}$ are positive definite
(in fact $\ge m$) frequencies corresponding to the modes 
${\bf k}$ and ${\bf l}$ in the out-region.
Clearly, the detector responds {\it only}\/ when the
Bogolubov coefffient~$\beta$ turns out to be non-zero 
(i.e. {\it only}\/ when particle production takes place).
However, it is evident that the transition probability rate 
of detector I have obtained above is {\it not}\/ proportional 
to the number of particles produced by the time-dependent
electric field background.

\section{In time-independent electric field\\ 
backgrounds}

Consider the vector potential
\begin{equation}
A^{\mu}=(A(x), 0, 0,0),\label{eqn:tine}
\end{equation}
where $A(x)$ is an arbitrary function of $x$.
Such a vector potential gives rise to a time-independent 
electric field along the $x$-direction given by ${\bf E}
=-(dA/dx)\,{\hat {\bf x}}$.
In such a case, the modes of the quantum field~${\hat \Phi}$ 
can be decomposed as follows:
\begin{equation}
u_{\omega\bf k_{\perp}}(t,{\bf x})
=e^{-i\omega t}\, f_{\omega{\bf k}_{\perp}}(x)\, 
e^{i{\bf k}_{\perp}\cdot{\bf x}_{\perp}},
\label{eqn:modetine}
\end{equation}
where ${\bf k}_{\perp}$ is the wave vector along the 
perpendicular direction.
Due to lack of time dependence, the Bogolubov coefficient 
$\beta$ relating these modes at two different times is 
trivially zero.
Though the Bogolubov coefficient $\beta$ is zero, particle 
production takes place in such backgrounds due to a totally 
different phenomenon.
It is well-known that if the depth of the potential $[qA(x)]$
is greater than $(2m)$, then the corresponding electric
field will produce particles due to Klein paradox (see 
Ref.~\cite{fulling89} and references therein). 
It is then interesting to examine whether an inertial detector 
in a time-independent electric field background will respond 
under the same condition.

Consider a detector that is stationed at a particular point.
It is easy to see from the form of the 
modes~(\ref{eqn:modetine}) that the transition 
amplitude~${\cal A}^{*}({\cal E})$ of such a 
detector will be proportional to a delta function 
as in the case of an inertial detector in the 
Minkowski vacuum (cf.~Eq.~(\ref{eqn:minkampnld})).
But, unlike the Minkowski case wherein the definition of
positive frequency modes match the definition of positive
norm modes, in a time-independent electric field background 
there exist negative frequency modes which have a positive 
norm whenever the depth of the potential $[qA(x)]$ is greater
than $(2m)$.
In other words, when Klein paradox occurs in an electric 
field background, $\omega_{k}$ and $\omega_{l}$ appearing 
in the argument of the delta function in 
Eq.~(\ref{eqn:minkampnld}) can be negative and, hence, 
there exists a range of values of these two quantities 
for which this argument can be zero.
These modes excite the detector as a result of which the 
response of an inertial detector proves to be non-zero in 
such a background. 

I shall now show (for the special case of the step potential) 
as to how there exist negative frequency modes which have a 
positive norm when the depth of the potential is greater
than $(2m)$.
In order to show that, let me evaluate the norm of the 
mode $u_{\omega{\bf k}_{\perp}}(t,{\bf x})$.
On substituting the mode~(\ref{eqn:modetine}) and the 
vector potential~(\ref{eqn:tine}) in the scalar 
product~(\ref{eqn:sclrprdct}), I obtain that
\begin{equation}
(u_{\omega{\bf k}_{\perp}},
u_{\omega{\bf k}_{\perp}})
= 2\, (2\pi)^2\, \delta^{(2)}(0) 
\int\limits_{-\infty}^{\infty} dx\,
\left[\omega-qA(x)\right]\,
\vert f_{\omega{\bf k}_{\perp}}(x)\vert^2.
\label{eqn:sp}
\end{equation} 
Let me now assume that $A(x)=-\left[\Theta(x)\,V\right]$, 
where $\Theta(x)$ is the step-function and $V$ is a constant. 
For such a case, the function $f_{\omega{\bf k}_{\perp}}$ 
is given by
\begin{equation}
f_{\omega{\bf k}_{\perp}}(x)
=\Theta(-x)\left(e^{ik_L x} + R_{\omega{\bf k}_{\perp}}\, 
e^{-ik_L x}\right)
+ \Theta(x)\,T_{\omega{\bf k}_{\perp}}\, e^{ik_R x},
\end{equation}
where 
\begin{equation}
k_R = \left[(\omega+qV)^2-\vert{\bf k}_{\perp}\vert^2
-m^2\right]^{1/2}\quad{\rm and}\quad
k_L = \left[\omega^2-\vert{\bf k}_{\perp}\vert^2
-m^2\right]^{1/2}.
\end{equation}
The quantities $R_{\omega{\bf k}_{\perp}}$ and 
$T_{\omega{\bf k}_{\perp}}$ are the usual reflection 
and tunnelling amplitudes.
They are given by the expressions
\begin{equation}
R_{\omega{\bf k}_{\perp}}
=\left(\frac{k_L-k_R}{k_L+k_R}\right)\quad
{\rm and}
\quad T_{\omega{\bf k}_{\perp}}
=\left(\frac{2k_L}{k_L+k_R}\right).
\end{equation}
If I now assume that $k_R$ and $k_L$ are real quantities, 
then, for the case of the step potential I am considering 
here, the scalar product~(\ref{eqn:sp}) is given by
\begin{equation}
(u_{\omega{\bf k}_{\perp}},
u_{\omega{\bf k}_{\perp}})
=(2\pi)^3\, \delta^{(3)}(0)\, 
\left[\omega \left(1+R_{\omega{\bf k}_{\perp}}^2\right)
+(\omega+qV)\, T_{\omega{\bf k}_{\perp}}^2\right].
\label{eqn:sp1}
\end{equation} 
Let me now set ${\bf k}_{\perp}=0$.
Also, let me assume that $\omega =-(m+\varepsilon)$ and 
$(qV)=(2m+\varepsilon)$, where $\varepsilon$ is a positive 
definite quantity.
For such a case, $R_{\omega0}=1$, $T_{\omega0}=2$ and
the scalar product~(\ref{eqn:sp1}) reduces to
\begin{equation}
(u_{\omega0},u_{\omega0})
=2\, (m-\varepsilon)\; (2\pi)^3\;  \delta^{(3)}(0)
\end{equation}
which is a positive definite quantity if I choose 
$\varepsilon$ to be smaller than~$m$. 
I have thus shown that there exist negative frequency modes
(i.e. modes with $\omega\le -m$) which have a positive norm. 
Moreover, this occurs {\it only}\/ when $(qV)$ is greater than~$(2m)$ 
(note that $(qV)=(2m+\varepsilon)$) which is exactly the condition 
under which Klein paradox is expected to arise.
As I have discussed in the last paragraph, it is this feature 
of the Klein paradox that is responsible for exciting the 
detector.

\section{In time-independent magnetic field\\ 
backgrounds}

A time-independent magnetic field background can be described 
by the vector potential
\begin{equation}
A^{\mu}=(0, 0, A(x), 0),\label{eqn:vecmag}
\end{equation}
where $A(x)$ is an arbitrary function of $x$.
This vector potential gives rise to the magnetic field 
${\bf B}= (dA/dx)\, {\hat {\bf z}}$, where ${\hat {\bf z}}$ 
is the unit vector along the positive $z$-axis.
It can be shown that the effective Lagrangian corresponding 
to such a time-independent magnetic field background does
{\it not}\/ have an imaginary part which then implies that such
backgrounds do {\it not}\/ produce particles~\cite{sriram96}.

The modes of the quantum field~${\hat \Phi}$ in a 
time-independent magnetic field background can be 
decomposed exactly as I did in Eq.~(\ref{eqn:modetine}) 
in the case of the time-independent electric field background.
Hence, the transition amplitude ${\cal A}^*({\cal E})$ 
of an inertial detector in a time-independent magnetic 
field background will also be proportional to a delta 
function as in Eq.~(\ref{eqn:minkampnld}). 
However, on substituting the mode~(\ref{eqn:modetine}) 
and the vector potential~(\ref{eqn:vecmag}) in the scalar 
product~(\ref{eqn:sclrprdct}), I find that
\begin{equation}
(u_{\omega{\bf k}_{\perp}},u_{\omega{\bf k}_{\perp}})
=\left(2\omega\right)\, (2\pi)^2\, \delta^{(2)}(0)\,
\int\limits_{-\infty}^{\infty} dx\,
\vert f_{\omega{\bf k}_{\perp}}(x)\vert^2
\end{equation} 
which is clearly a positive definite quantity whenever 
$\omega\ge m$.
In other words, unlike the case of the time-independent 
electric field background, in a time independent magnetic 
field background, the definition of positive frequency modes
{\it always}\/ match the definition of positive norm modes.
Therefore, as in the case of an inertial detector in the
Minkowski vacuum, an inertial detector will {\it not}\/ respond 
in the vacuum state in a time-independent magnetic field 
background.

\section{Concluding remarks}

It is clear that, when the quantum field is in the vacuum 
state, the non-linearly coupled detector, while in {\it 
inertial}\/ motion, responds {\it only}\/ when the classical 
electromagnetic background produces particles.
However, as we have seen in the case of the time-dependent 
electric field background, the detector response does 
{\it not}\/ reflect the amount of the particles produced by 
the background.
This feature should not come as a surprise and, in fact, it 
can be attributed to the non-linearity of the interaction
Lagrangian~(\ref{eqn:giint}) for the following two reasons. 
Firstly, it is known that in a time-dependent gravitational 
background with asymptotically static domains, the response 
of the Unruh-DeWitt detector (which is coupled to the quantum 
field through a linear interaction) in the out-region {\it is}\/ 
proportional to the number of particles produced by the 
background (see Ref.~\cite{bd82}, pp.~57-59).
Secondly, it has been shown that the response of a detector 
that is coupled to the energy-momentum tensor of the quantum 
field (which is evidently a non-linear interaction) does {\it not}\/ 
reflect the particle content of the field~\cite{paddytp87}.
As I have discussed earlier, demanding gauge invariance 
naturally leads to non-linear interaction Lagrangians. 
Therefore, quite generically, we can expect that the 
response of detectors in classical electromagnetic 
backgrounds will {\it not}\/ be proportional to the 
amount of particles produced by the background.

\section*{Acknowledgments}

I would wish to thank Prof.~Jacob D.~Bekenstein and 
Prof.~T.~Padmanabhan for discussions.
This work was supported in part by a grant from the Israel 
Science Foundation established by the Israel Academy of 
Sciences.

\end{document}